\documentclass[%
 reprint,
 amsmath,amssymb,
 aps,
]{revtex4-2}

\usepackage{graphicx}
\usepackage{dcolumn}
\usepackage{bm}
\usepackage{hyperref}
\begin{document}
\title{Asymmetric reformulation of draw rules in chess and its implications for game theory: Repetition as loss for White}
\author{Chong Qi}
 \affiliation{Department of Physics, KTH Royal Institute of Technology, AlbaNova University Centre, 106 91 Stockholm, Sweden}
\date{\today}
\begin{abstract}
Repetition-based draw rules in deterministic games like chess ensure termination but introduce strategic artifacts, allowing players to enforce draws independent of positional value. We propose an asymmetric modification: threefold repetition results in a loss for White if it is responsible for initiating it. This rule directly targets the persistent first-move advantage and removes low-effort draw strategies available to White. The new rule is expected to reduce draw rates, re-balance first-move advantage, and promote exploration in both human and artificial play. We outline a computational framework with existing and newly designed neural-network chess engines for the empirical validation of  the proposal and analyze it from the perspectives of game theory and graph dynamics.
\end{abstract}
\maketitle
\section{Introduction}

Deterministic, perfect-information games like chess serve as foundational models in mathematics and artificial intelligence and have long been used as a benchmark for search algorithms, evaluation functions, and reinforcement learning systems.
A central issue in such games is the handling of repeated positions. The \emph{threefold repetition rule} declares a draw when the same position occurs three times, ensuring termination in a finite state space. However, this rule also introduces unintended strategic consequences: players may deliberately steer the game into repetition to avoid loss. In addition, there is a persistent first-move advantage that is not fully compensated under current rules. Empirical studies of high-level chess games show that White consistently achieves a higher expected score than Black, typically around 54--56\% \cite{wiki_chess_advantage_2026}.
In reality, a master level White can nearly always force a draw, leading to the \emph{safe draw strategies}.

Ref. \cite{Tomasev2020}, co-authored by former world chess champion, recognizes that modern professional chess faces challenges from a high percentage of draws and many games that end in preparation phase without fighting. This trend is exacerbated by an increasing reliance on engine theory and extensive opening preparation, which can limit the creative scope of the game. The study explored 9 ways of rule changes and whether atomic rule changes can revitalize the game by introducing novel strategic and tactical patterns.

Ref. \cite{Barthelemy2025} provides a systematic computational analysis of all the possible starting positions in the so-called Fischer Random Chess. A striking feature it reveals is that the first-move advantage is a persistent and structural characteristic across all chess setups, not just the standard starting position. This finding suggests that the inherent edge provided by moving first is an inescapable feature of the game’s geometry and rules, with a mean initial evaluation of $+0.33 \pm 0.12$ pawns regardless of the initial piece displacement.

The standard repetition rule produces outcomes that do not necessarily reflect the intrinsic value of positions. A player in a worse position may force a draw by constructing a repetition sequence, leading to what can be described as \emph{artificial draws}. The availability of repetition reduces incentives for
risk-taking and exploration or
 innovative strategic planning.
Instead, it encourages defensive and cyclic strategies that prioritize safety over progress, which can make the game notoriously boring.
 These facts raise a fundamental question: should rules merely prevent infinite play, or should they also shape incentives toward meaningful progress?

Therefore, we propose a minimal modification to the rules that reduces White’s first-move advantage while simultaneously discouraging overly defensive play and promoting more dynamic, spectator-friendly games, referred to as asymmetric repetition rule (ARR). 
The modified rule is defined as follows:
\begin{quote}
If a threefold repetition occurs and White is responsible for initiating the repetition, the game is scored as a loss for White.
\end{quote}

The key features of the ARR are as follows. First, while formally asymmetric, the rule acts to counterbalance the inherent asymmetry of the first move, thereby driving the game toward greater practical equality. Second, repetition is no longer a risk-free drawing mechanism for White. Crucially, this change restores strategic responsibility to White, who can no longer rely on passive play to secure an artificial draw through repetition. Third, Black retains repetition as a legitimate defensive resource, preserving the ability to secure a draw in objectively inferior positions inherent to second play, while retaining practical opportunities to stabilize the position or reverse the evaluation. Finally, the rule preserves the historical record of games, altering only the interpretation of outcomes under repeated positions, and introduces little to no learning cost for players adapting to the new rule.

\section{The draw problem in chess}

\begin{figure*}[!ht]
    \centering
    \includegraphics[width=0.75\linewidth]{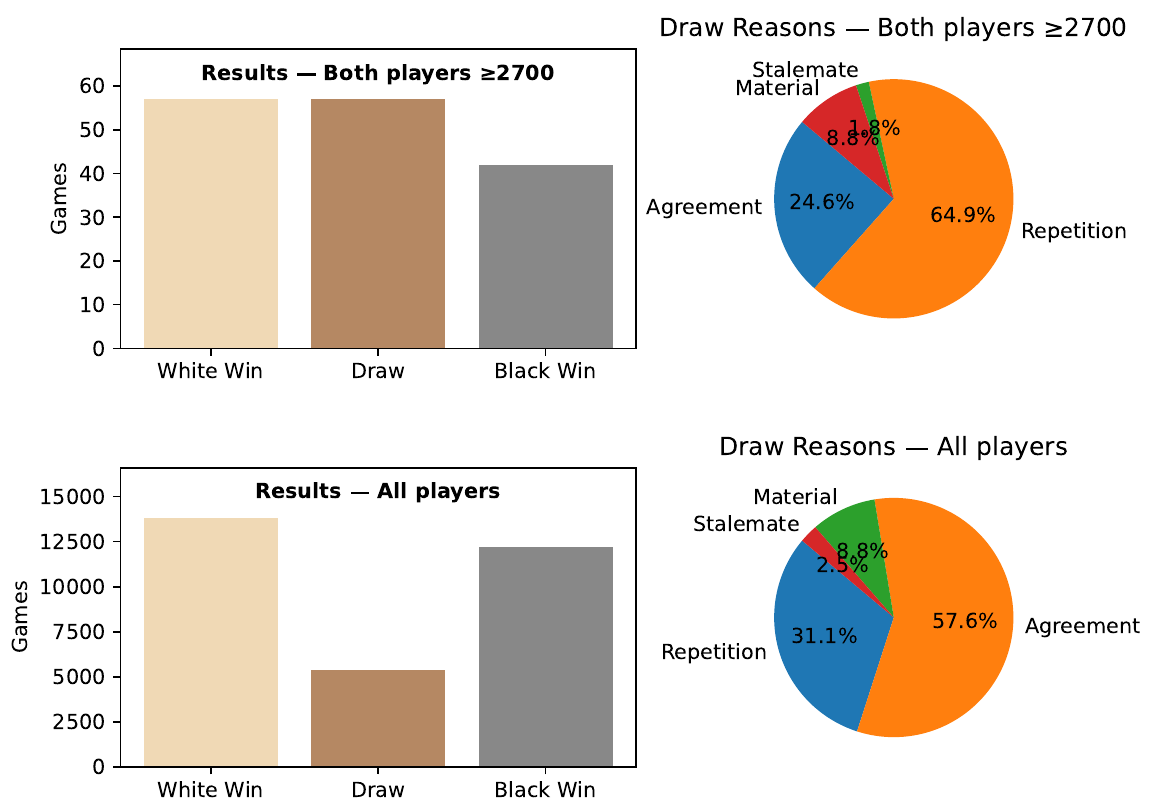}
    \caption{Game outcome distributions and draw reason breakdowns for games retrieved from the latest issues of TWIC~\cite{twic}, comprising international tournament, including the 2026 Candidates Tournament,
 and competitive club games collected across a broad range of playing levels and events worldwide played in March, 2026. }
    \label{fig:placeholder}
\end{figure*}
In general,  the ``draw problem" is a central theme of competitive balance in the modern era of high-level board games. While some traditional board games use distinct mechanical ``anchors" to ensure decisive outcomes~\footnote{In Chinese chess (Xiangqi), which shows high similarity to chess, repetition is a loss: you lose the game if you perpetually check or  chase a piece and refuse to change. Shogi (Japanese chess) has the lowest draw rate of any major board game due to the so-called drop rule where captured pieces can be placed back on the board as your own.
In the game of Go (Weiqi), which is mathematically distinct due to the uniform nature of its pieces, the Komi compensation is used to balance the first-move advantage by awarding extra points to the second player, leading to practically zero draw.
}, chess is currently grappling with an extremely high draw rate, up to around 65-75\% at the elite level and 95\% for top chess engines. 
In top-tier classical (long duration) chess, the most common result is a draw. As players use powerful chess engines for preparation, the margin for error has shrunk, leading to high-accuracy games where neither side can gain a decisive edge. In the 2018 chess world championship match (Carlsen-Caruana), all 12 classical games ended in draws. In the 2021 championship match (Carlsen-Nepomniachtchi), there were five consecutive draws until a decisive game which lasted 136 moves (nearly 8 hours), becoming the longest game in championship history. The latest 2024 championship match (Gukesh-Ding) was a dramatic departure featuring five decisive games, mostly due to the psychological factors of both competitors rather than a high-level of aggressive play. 

Due to the high frequency of draws,  the ``draw death" is becoming a significant concern. There have been ongoing discussions to abandon classical games in favor of faster formats, to ensure more decisive and spectator-friendly results.

Nevertheless, draws in chess arise from several distinct causes, both rule-based and practical. One common outcome in high-level play is a draw by mutual agreement, where both players consent to halt the game, typically when the position is assessed as balanced or when further play is deemed unlikely to produce a decisive result. Other situations such as stalemate, the fifty-move rule, and insufficient mating material also lead to a draw. Among these, the most dramatic and arguably most controversial is the draw by threefold repetition, which occurs when the same position arises three times with the same side to move. In practice this most frequently manifests as perpetual check, where one side delivers an unending sequence of checks that the opponent cannot escape, but it also opens the door to passive play and artificial draws in which the disadvantaged side exploits the rule to escape an otherwise losing position. An extreme real-world illustration of passive drawing behavior is the deliberate use of threefold repetition at a very early stage of the game, bypassing meaningful play entirely. Such occurrences have been noted in the literature~\cite{Tomasev2020} and have arisen in practice at the highest level of professional competition~\cite{FIDE2023dubov}.

It is precisely this last aspect — that a strategically unmotivated player may secure half a point not through resourceful defense or attack but through deliberate repetition — that has motivated proposals to revise or replace the threefold repetition rule.

Several databases are regularly updated with games from both online platforms and over-the-board tournaments. The Lichess open database records approximately 90 million games per month \cite{lichess_data}, while This Week in Chess (TWIC) collects around 35,000 tournament games over the same period \cite{twic}. A statistical overview of the TWIC dataset is shown in Fig.~\ref{fig:placeholder}. The upper panels display results for games in which both players hold a FIDE rating of at least 2700, while the lower panels show the corresponding distributions for the full dataset without rating restriction. The left panels show the outcome frequencies — White win, Black win, and draw — and the right panels show the composition of drawn games, distinguishing agreement or technical draws, draws by insufficient material, draws by stalemate, and threefold repetition. Notably, the draw rate in the high-rated class is substantially lower than typically observed in elite play, partly attributable to the inclusion of the 2026 Candidates Tournament and the preparation games of some of its players, where one may expect to adopt an aggressive, result-oriented approach.
In general, the draw rate is usually lower for younger and relatively lower rated players.

The outcome distributions shown in Fig.~\ref{fig:placeholder} are instructive from a game-theory perspective. Game theory studies how rational agents make decisions when their outcomes depend not only on their own choices but also on the choices of others. In the context of chess, each player selects a strategy, a plan of actions for every possible position, with the goal of maximizing their individual payoff, which is determined not by the quality of the moves alone but by the final result and its broader implications for standings and qualification. This perspective offers a natural explanation for the generally elevated draw rate in professional chess, where players often weigh safety above winning, as well as for the notably lower draw rate observed in the recent dataset, where a win carries substantially greater value than a draw, as is the case in the 2026 Candidates Tournament. Under such incentive structures, rational players accept greater risk and pursue decisive outcomes even from objectively balanced positions.

Several proposals have been put forward to modify the rules of chess in ways that alter the payoff structure and encourage more ambitious, risk-taking play~\cite{Tomasev2020}. Most of these, however, intervene at a fundamental level and alter the essential character of the game. The ARR proposed here is distinguished from such approaches in that it leaves the rules of play entirely intact (the same moves remain legal, the same positions can arise) while introducing a targeted asymmetry in the consequence of threefold repetition that shifts the incentive structure away from passive drawing strategies. It is, to our knowledge, the only proposed modification that changes the game's payoff without changing its nature.

A natural alternative would be to penalize threefold repetition symmetrically regardless of color. However, our simulations demonstrate that such a symmetric rule significantly amplifies the first-move advantage of White, as Black is deprived of the repetition escape that currently serves as a crucial defensive resource. This would alter the fundamental balance of the game and the nature of the skills it rewards. The ARR avoids this pitfall by preserving the repetition escape exclusively for Black, without disturbing the strategic equilibrium between the two sides.

\section{Numerical evaluation of the asymmetric repetition rule}
The proposed ARR has several important consequences that must be systematically examined. First, it directly alters the payoff structure of the game by penalizing repetition completed by White, thereby discouraging passive drawing strategies and forcing more active play from the first player. Second, it introduces an asymmetry that counteracts the natural first-move advantage of White, raising the question of whether the game approaches a more balanced and symmetric equilibrium under optimal play. Third, it preserves repetition as a defensive resource for Black, which may strengthen defensive strategies and shift the burden of decisive play more heavily onto White.

A careful numerical evaluation must therefore address several key questions. Does the rule produce a meaningful increase in decisive games without distorting the overall balance between White and Black? Does it disproportionately favor Black to the point of introducing a second-move advantage? Does it influence opening theory?
Crucially, the rule should not require players to fundamentally relearn the game. In other words, the vast majority of chess knowledge, opening theory, tactical patterns, and endgame technique must remain valid and applicable. 

To start with, one can safely state that, from a learning perspective, the rule introduces only a localized modification to the evaluation landscape, suggesting that both human players and neural networks can adapt with minimal retraining while still exhibiting measurable shifts in behavior.

To evaluate the impact of the ARR, we design a three-level evaluation framework aimed at capturing both immediate and long-term effects on strategy and theoretical balance. At the first level, we employ a strong reference engine, Stockfish, to simulate games under both the standard rules and the modified repetition rule, enabling a direct comparison of outcomes, draw rates, and practical playing dynamics. This provides a baseline measure of how the rule alters high-level engine behavior without retraining.
It also helps to quantify the proportion of draws in chess that rely on White-initiated threefold repetition under standard engine play.
At the second level, we train a neural network algorithm to estimate the initial advantage of White. The network is trained on a combination of large-scale historical datasets and self-play games. This step is essential to quantify whether the modified rule successfully reduces or eliminates the inherent first-move advantage.
At the third level, we build a neural-network chess engine by fine-tuning its evaluation under the modified rule to study how optimal play adapts in long term strategy when repetition carries asymmetric consequences. This allows us to analyze not only static evaluation shifts, but also changes in strategic incentives and learned behavior.

\subsection{Stockfish Self-Play Evaluation}

\begin{table*}[!ht]
\centering
\caption{The seven experimental setups of Stockfish self-play under standard rules and ARR. Depth controls playing strength 
and contempt controls draw-avoidance tendency. Positive contempt 
encourages the engine to avoid draws; negative contempt makes it more 
willing to accept them.}
\label{tab:setups}
\begin{tabular}{clcccc}
\hline
Setup & Description & Depth (W) & Depth (B) & Contempt (W) & Contempt (B) \\
\hline
A & Equal strength, neutral       & $d$   & $d$ &   0  &   0  \\
B & Equal strength, contempt      & $d$   & $d$ & +50  & $-$50 \\
C & White stronger, neutral       & $d+2$ & $d$ &   0  &   0  \\
D & White stronger, aggressive    & $d+2$ & $d$ & +50  & $-$50 \\
E & White weaker, fighting        & $d-2$ & $d$ & +50  &   0  \\
F & White weaker, neutral         & $d-2$ & $d$ &   0  &   0  \\
G & White weaker, defensive       & $d-2$ & $d$ & $-$50 &   0  \\
\hline
\end{tabular}
\end{table*}
To evaluate the ARR numerically, we employ Stockfish~\cite{stockfish} in a controlled 
self-play framework. Each game begins randomly from one of 100 selected major opening systems. The experiment is repeated across seven distinct setups, labeled A 
through G in Table \ref{tab:setups}, that systematically vary two engine parameters: search depth, 
which controls playing strength, and contempt, which controls the 
engine's willingness to accept draws. The base search depth is denoted 
$d$; stronger play corresponds to depth $d+2$ and weaker play to $d-2$.
For each experimental setup, two independent runs are conducted under the standard rule and ARR:

{Run~1} plays under the standard rules, extracts  the standard result, serving as the baseline and re-classifies the same games post hoc under the ARR: 
any repetition completed by White is reclassified as a Black win, while 
repetitions completed by Black remain draws. This naive output models the 
effect of the rule with zero adaptation by White. It reveals the very high percentage of threefold repetitions initialized by White. 

{Run~2} plays in which White is 
\textit{aware} of the ARR and actively avoids completing repetition 
cycles. This is implemented by filtering White's move choices via the 
Stockfish {searchmoves} option: at each move, any move that 
would cause the position to occur for the third time is excluded from 
White's search, forcing Stockfish to select the best available 
non-repeating continuation. 

\begin{figure}
    \centering
    \includegraphics[width=0.995\linewidth]{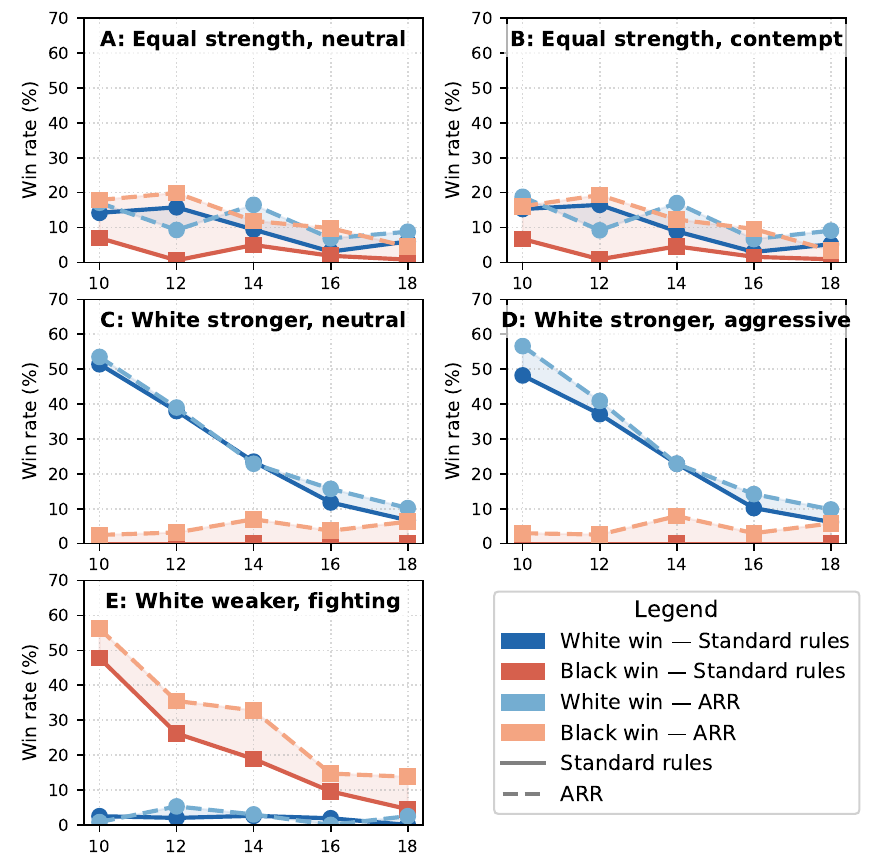}
    \caption{Win rates for White and Black as a function of Stockfish search depth across the different experimental setups A--E, comparing the standard threefold repetition rule (solid lines) with the asymmetric repetition rule (dashed lines). In each panel, the shaded band between the solid and dashed lines of the same color illustrates the net effect of the ARR relative to the standard rules.}
    \label{fig:abcde}
\end{figure}

Setup~A serves as the symmetric baseline, with both sides playing at equal strength and no draw preference, while Setup~B introduces contempt asymmetry — White avoiding draws, Black welcoming them — while keeping playing strength equal. In both cases, the ARR produces a notably larger enhancement in Black's winning rate than in White's, which is the expected and desired outcome: under equal strength conditions, the rule's asymmetric penalty on White-completed repetitions translates directly into additional winning opportunities for Black, without substantially disrupting White's own winning chances. The overall draw rate decreases meaningfully in both setups, confirming that the ARR successfully destabilizes the passive drawing equilibrium even when neither side holds a material or positional advantage.

Setups~C and D model a scenario in which White is objectively stronger, with a two-ply depth advantage over Black. White's superiority is clearly reflected in the outcome distribution, with White winning rates substantially exceeding those of Black across all depths. Under the ARR, however, this dominance is partially offset: by penalizing White-completed repetitions, the rule removes a key mechanism through which the stronger White engine would otherwise consolidate its advantage and force a draw from a winning position. This gives Black a more meaningful opportunity to contest the game, and the Black winning rate rises noticeably under the ARR. This is precisely the corrective effect the ARR is designed to produce: preserving White's legitimate winning chances while reducing the ability to exploit the repetition rule as a risk-free drawing tool.

Setups~E, F, and G model a weaker White, examining whether Black can exploit the ARR more readily when already in a superior position, and how White's contempt  modulates this effect. Since the three setups yield qualitatively similar results, only Setup~E is shown for brevity. The stronger Black engine dominates under the standard rules, particularly at lower search depths where the strength differential is most pronounced. As expected, the dominance is enhanced by ARR since White cannot escape with repetition. Nevertheless, a notable finding is that the ARR consistently enhances White's winning chances even in this disadvantaged configuration — an initially counterintuitive result that can be understood as follows: by being forced to avoid repetition cycles, the weaker White is compelled to seek active, complex continuations rather than retreating into drawing lines, and these complications occasionally generate winning opportunities that would not have arisen under passive play. This suggests that the ARR does not simply penalize White, but restructures the incentive landscape in a way that can benefit both sides by driving the game toward more decisive and dynamic positions.

Together, the seven setups span a representative range of competitive scenarios and allow the effect of the ARR to be assessed across varying strength differentials and risk dispositions. The results consistently show that the ARR reduces the draw rate in engine self-play, and remarkably, it can enhance the winning rate for both White and Black simultaneously — a signature of more decisive, fighting chess in which fewer games are resolved by repetition and more by genuine play.  In practice, however, engine self-play captures only the purely rational response to the new incentive structure. Human players bring additional dimensions that engines do not — psychological pressure and the cognitive burden of navigating a rule that asymmetrically penalizes one side. These factors are likely to amplify the effect of the ARR beyond what is observed in the simulations. A definitive assessment of the rule's practical impact therefore requires testing in real over-the-board games.

\subsection{Neural network evaluation of the initial advantage and experimental fine-tuning}

The initial evaluation is a numerical score that represents the theoretical advantage held by  White due to the first-move advantage \cite{wiki_chess_advantage_2026,Ribeiro2013-xh}. Modern engines like Stockfish typically evaluate this starting position between +0.22 and +0.25, which reflects the statistical probability of a win. This score is measured in pawns instead of win probability, where an evaluation of +0.2  indicates an advantage equivalent to one fifth extra pawn. The transformation can be roughly expressed
as
$$\text{Evaluation} \approx (\text{win probability}) \times 4$$
and
$$\text{win probability} = \tanh(\text{Evaluation} )$$
which means an advantage of 0.22 pawns corresponds to a win probability of approximately 0.055 (or 5.5\%), while an evaluation of 4 pawns represents absolute certainty of a win.

To evaluate the impact of the ARR on the initial advantage, we developed a two-stage training pipeline based on a ResNet-style NNUE (Efficiently Updatable Neural Network) architecture. We started by training the ResNet against Stockfish self-plays up to depth 20, which however is very slow for us. Therefore we decided to use an existing massive dataset of approximately 362 million positions from Lichess database \cite{lichess_evals_2026}, which are also pre-evaluated with Stockfish.
In the first stage, the network is trained to predict Stockfish-evaluated outcomes from these historical positions. This provides a strong baseline model of standard chess evaluation.
In the second stage, we construct a modified network by fine-tuning the converged weights under the modified repetition rule. 
In this way, the modified network preserves general chess knowledge while adapting specifically to the altered strategic incentives introduced by the rule change.
During fine-tuning, move selection is guided by playing with Stockfish, but White's repetition-inducing moves are filtered to simulate the new penalty mechanism. In principle, this adjustment should be relatively easy for a human player to adapt to, as it primarily affects a narrow subset of strategic situations rather than the full evaluation function.

The initial untrained NNUE-style model yields an average evaluation of +0.29; however, full convergence has not yet been achieved for the standard network or the modified one, primarily due to computational constraints in reproducing the high-depth Stockfish evaluations (depth 30–36) used in the Lichess dataset.

\subsection{Self-play chess engine under the modified rule}
One may argue that an intrinsic bias exists in standard engines such as Stockfish, which may implicitly account for the possibility of threefold repetition in future search steps during evaluation. We consider this effect to be minor.
Anyhow, to quantify the impact of the modified repetition rule on White’s first-move advantage, we developed a self-play chess engine in which any repetition initiated by White is scored as a loss ($-1.0$), thereby discouraging “escape” draws. The engine is built again around an NNUE-style architecture. This neural network is used in Stockfish while AlphaZero used a different deep residual convolutional networks combined with Monte Carlo tree search. Our implementation follows the NNUE paradigm and is implemented in TensorFlow.
The model is optimized for fast evaluation of board states, enabling large-scale self-play training exceeding two million games. Although the new engine remains relatively premature compared to Stockfish, the results are already quite appealing and promising.
Under this modified rule, White’s initial advantage is substantially reduced, from approximately 0.22 pawns to just 0.009–0.013, indicating an almost perfectly balanced position.

In addition, we analyzed approximately 10,000 pre-repetition positions in which a network trained under standard rules elects to repeat. The standard-trained network evaluates these positions at $+5.1 \pm 13.8 $ centipawns (from White’s perspective), suggesting that repetitions are typically initiated from equal or slightly unfavorable positions.
When the same positions are evaluated by the modified-rule network, the average score decreases to $+3.9 \pm 12.3$ centipawns, corresponding to a mean difference of -1.2 centipawns. This result is expected because the modified rule makes such positions more difficult for White to hold. But the real difference is extremely small.

\section{Graph-theory formulation and equilibrium structure}
Consider two even players for a game: if both adopt aggressive, risk-taking approaches, each faces a meaningful probability of losing; if both play solidly and avoid unnecessary complications, a draw is the likely outcome. The result is a Nash equilibrium in passive play: both players choose low-risk strategies, neither can improve by deviating unilaterally, and the game ends in a draw.
Nash equilibrium is a central concept in game theory, which is defined as a configuration of strategies in which no player can improve their individual outcome by unilaterally changing their own strategy, assuming all other players maintain theirs. It identifies stable  situations that rational players will naturally gravitate toward and have no incentive to deviate from. 
Theoretically, this equilibrium is a direct and predictable consequence of game and may be inevitable as chess engines get stronger.

The threefold repetition rule allows another kind of equilibrium: a player who is objectively worse  to secure an artificial draw that would not be achievable on the merits of the position alone, effectively preventing the stronger player from converting a winning advantage. Both players know this option exists at all times. 
In other words, the
equilibrium behavior in chess is highly sensitive to the payoff assigned to repetition. By altering repetition from a neutral outcome to a penalized outcome, one removes a safe equilibrium path and forces players into risk-dominant strategies.

The ARR intervenes precisely at this point. By converting threefold repetition completed by White from a draw into a loss, it removes the previously guaranteed fallback and reshapes the payoff structure in a way that destabilizes the passive equilibrium not by forcing players to play differently. This shifts the equilibrium toward risk-taking behavior and creates asymmetric strategic incentives.

Chess can be formally represented as a finite directed graph in which the vertices correspond to legal board positions and the directed edges correspond to legal moves between them. Crucially it contains cycles which turn the game back to a previously visited position. 
Under the standard rules, the threefold repetition provision maps every cycle onto a terminal node with a neutral payoff: a draw. In graph-theoretic terms, this transforms cycles into absorbing equilibrium states: once a repeated position is available, the player who can claim it faces no further uncertainty, and will go for it. 

The ARR does not remove cycles from the graph, where the same positions can still recur, and the same sequences of moves remain legal. What changes is the payoff assigned to cycles depending on which player completes them. For White, a cycle is no longer a neutral absorbing state but a losing terminal node. This asymmetric reassignment of cycle payoffs has a precise game-theoretic consequence as discussed above: the absorbing draw equilibrium that previously supported passive play by White is destabilized, and the optimal strategy profile shifts toward risk-dominant play. 
In other words, it asymmetrically removes these absorbing states for one player and forces optimal play toward decisive continuations. The ARR thus modifies the equilibrium structure of chess at the level of the graph's payoff function, without altering its topology or nature of the game.

\section{Conclusion and discussion}
The proposed ARR modification is expected to reduce White’s first-move advantage, discourage passive play and promotes more decisive game-play, particularly in high-level and human games where repetition is often used as a safety mechanism. This may lead to more decisive and spectator-friendly outcomes. Importantly, ARR preserves the full historical record of chess games, altering only the interpretation of outcomes in repeated positions. The learning cost associated with the rule appears to be minimal:  human players are expected to adapt relatively quickly, as the modification affects a localized subset of positions rather than the overall structure of the game.
Nevertheless, it also introduces several conceptual and practical considerations that warrant careful examination.

A primary concern is the introduction of asymmetry in the treatment of repetition even though one can argue that the game itself is not symmetric due to the existence of move order and persistence of first move advantage. Repetition, previously a valid equilibrium strategy for both players, becomes a dominated strategy for White: White may be forced to avoid repetition, potentially leading to suboptimal or even losing positions while Black may be more keen to steer the game into repetition. 
Nevertheless, the concept of asymmetric incentives in chess is not without precedent. In the so-called Armageddon tiebreak, White is required to win to claim victory, while a draw is sufficient for Black. However, it can be very challenging there to balance the competing advantages of each side.

At the same time, the rule enhances the defensive role of Black by preserving repetition as a fallback mechanism. This ensures that objectively inferior positions due to second play can be stabilized. From this perspective, the rule does not introduce asymmetry so much as correct an existing one: the standard threefold repetition rule disproportionately benefits White by providing a risk-free escape from positions where White holds the initiative but cannot convert it into a win. White may be compelled to take on additional risk to win the game. On the other hand, one can argue that that additional risk should have been part of the game, as White sets the tone of the game and has the responsibility to pursue a decisive outcome, which is artificially relaxed by the threefold repetition rule.

From a game-theoretic perspective, the modification alters the payoff structure rather than move dynamics, therefore preserving the nature of the game. Under standard rule, threefold repetition represents a neutral equilibrium outcome that attracts both sides,  irrespective of their ability to make progress. Under the proposed rule, this equilibrium is broken: repetition initiated by White becomes a losing outcome, while repetition initiated by Black remains a draw. This constitutes a deliberate trade-off: the elimination of low-effort drawing strategies in exchange for a more decisive and dynamic game.

From a practical standpoint, the ARR is also likely to increase the variance of game outcomes, at least in the short term. This effect arises not only from the altered payoff structure but also from psychological factors influencing player behavior, as players with White are forced to avoid stable repetition and to navigate for a win in more complex and potentially unbalanced positions.
and to seek winning chances in more complex and potentially unbalanced positions. At the same time, Black is incentivized to take greater risks, as the preserved repetition escape reduces the structural weight of the second-move disadvantage. This is exactly what is needed for chess to escape the draw death.

\section*{Acknowledgment}
I thank Jan Dufek for allowing me to use the CPU and GPU nodes of the MEGARAPTOR cluster at KTH during its testing phase.
\bibliography{ref}
\end{document}